\documentclass[cameraready]{Interspeech}
\usepackage{algorithm}
\usepackage{algorithmic}
\usepackage{amsmath}
\usepackage{stfloats}
\usepackage{caption}
\usepackage{tabularx}
\usepackage{stfloats}
\usepackage{amssymb}
\usepackage[figuresright]{rotating}
\usepackage{lineno}
\usepackage{hyperref}
\usepackage{bm}
\usepackage{subfigure}
\usepackage{multirow}
\usepackage{pifont}
\usepackage{mathrsfs}
\usepackage{makecell}
\usepackage{color}
\usepackage{array}
\usepackage{graphicx}
\usepackage{rotating}
\usepackage{amsfonts}

\title{Noisy Environment Adaptation of Neural Speech Codec via \\Focal Mask and Noise Feature Separation}

\author[affiliation={1}, orcid=0000-0002-1684-043X]{Shaokai}{Li}
\author[affiliation={1,2}, orcid=0000-0000-0000-1111, correspondingauthor]{Weiping}{Tu}
\author[affiliation={1,2}, orcid=0000-0003-3001-7957]{Yuhong}{Yang}

\address{
    $^1$ National Engineering Research Center for Multimedia Software (NERCMS), School of Computer Science, Wuhan University, China \\
    $^2$ Hubei Key Laboratory of Multimedia and Network Communication Engineering, China
}
\email{2025102110033@whu.edu.cn, tuweiping@whu.edu.cn, yangyuhong@whu.edu.cn}

\usepackage{comment}

\begin{document}

\maketitle

\begin{abstract}
Neural speech codec has attracted extensive attention for high-quality reconstruction at low-bitrate. However, real-world noise severely degrades its performance and hinders high-quality clean speech reconstruction. To tackle this problem, we propose FocalSE, a novel speech enhancement method that performs feature denoising, noise feature separation and noise recognition in the continuous embedding space of neural speech codecs. Specifically, we develop focal modulation-based compression and decompression to capture global context and local mutual information, and generate focal masks to recover clean feature embeddings. We then separate noise embeddings from noisy embeddings to improve denoising performance. Finally, we use ResNet1D-18 to recognize noise categories for better separation effectiveness. Extensive experiments on two standard datasets, LibriTTS and ESC50, demonstrate that our method outperforms state-of-the-art approaches under low-bitrate and low-SNR conditions.

\end{abstract}

\section{Introduction}

Neural speech codecs (NSC) aims to reconstruct high-quality speech signals at the lowest possible bitrates, thereby improving the storage efficiency of media communications \cite{kim2025neural}. In recent years, with the continuous iteration of large language models (LLMs), discrete token representation has become the core technical cornerstone of NSC \cite{cui2025recent}, and lots of state-of-the-art NSC methods have been proposed, such as SoundStream \cite{zeghidour2021soundstream}, DAC \cite{kumar2023high}, FocalCodec \cite{della2025focalcodec}, and so on \cite{mousavi2025discrete}. All these methods support bitrate control and achieve high-quality compression and reconstruction of speech signals.\par
However, in real-world scenarios, speech signals are often contaminated by various environmental noises \cite{zmolikova2023neural}, such as natural ambient sounds, urban noise, and human non-verbal noise. These noises will cause existing NSC methods to deviate from ideal signal generation conditions, leading to severe distortion and quality degradation of reconstructed signals. Fortunately, speech enhancement (SE) strategies can effectively tackle this issue, with the goal of mitigating noise interference and signal distortion in noisy environments \cite{zheng2023sixty}. Traditional SE methods directly process noisy speech to reconstruct clean signals \cite{jun2025snr, song2025lightweight}, and though effective to some extent, they suffer from low efficiency and cannot be seamlessly integrated with the NSC framework to satisfy low-bitrate constraints.\par
To address the above issues, many SE methods are designed to perform in the continuous embedding space of NSC. We roughly divide these NSC-based SE methods into two categories: clean token generators and clean embedding extractors. Clean token generators \cite{wang2024selm, yang2024genhancer, xue2024low, liu2025neural, yaogense} use self-supervised models to generate clean tokens from quantized noisy tokens for speech reconstruction. In contrast, clean embedding extractors \cite{li2025speech, mu2025continuous, chae2025towards, han2025dual, shetu2025gan} extract clean embeddings from encoded noisy embeddings by designing feature denoising strategies, followed by quantization or decoder. According to \cite{li2025speech, kammoun2025modeling}, in this paper, we focus on the clean embedding extractor approaches, which has higher inference efficiency and excellent speech reconstruction results. \par
Recently, several NSC-based SE methods have been proposed to improve the adaptability of NSC in noisy environments. Li et al. \cite{li2025speech} design a efficient SE module, which employs Transformer \cite{vaswani2017attention} to learn global masks in the low-dimensional space obtained by down-sampling feature channels, so as to recover clean embedding features.
Mu et al. \cite{mu2025continuous} improve the performance of SE by conducting progressive clean embedding extraction through two stages of encoding and quantization within the continuous embedding space of NSC. Chae et al. \cite{chae2025towards} integrate the SEMamba module \cite{WOS:001440556800041} and variable bitrate into the embedding space of NSC to enhance the signal reconstruction performance in noisy environments. Han et al. \cite{han2025dual} adopt the knowledge distillation strategy to extract clean embeding features during the quantization process of NSC. Shetu et al. \cite{shetu2025gan} leverage the masked multi-head attention mechanism in the encoding space to improve the performance of SE. Kammoun et al. \cite{kammoun2025modeling} conduct a detailed analysis of the signal reconstruction performance of token generation strategies and feature extraction strategies in the continuous embedding space of NSC. However, most existing NSC-based SE methods only focus on the clean target while neglecting the learning of noise components to be suppressed, which degrades the performance of speech reconstruction under low-bitrate and low-SNR conditions.\par
Different from existing works, in this paper, we propose a novel NSC-based SE method, named FocalSE. Our contributions are summarized as follows:
\begin{itemize}
    \item We design a novel feature mask strategy, named focal mask, which captures global contextual and local mutual information of speech signals by leveraging the focal modulation mechanism and Transformer blocks.
    \item We use focal mask to perform feature denoising, noise feature separation, and noise recognition in the continuous embedding space of NSC. These components can reinforce each other, enabling NSC to better adapt to noisy environments.
    \item Under various low-bitrate and low-SNR conditions, we conduct extensive experimental comparisons on speech reconstruction between the proposed method and state-of-the-art baselines, while also analysis the noise recognition performance of our method.
\end{itemize}
\begin{figure*}[ht]
\centering
\setlength{\abovedisplayskip}{0pt}
\setlength{\belowdisplayskip}{0pt}
\begin{minipage}{0.61\textwidth}
    \centering
    \subfigure[\footnotesize The framework of FocalSE.]{
        \includegraphics[width=\linewidth]{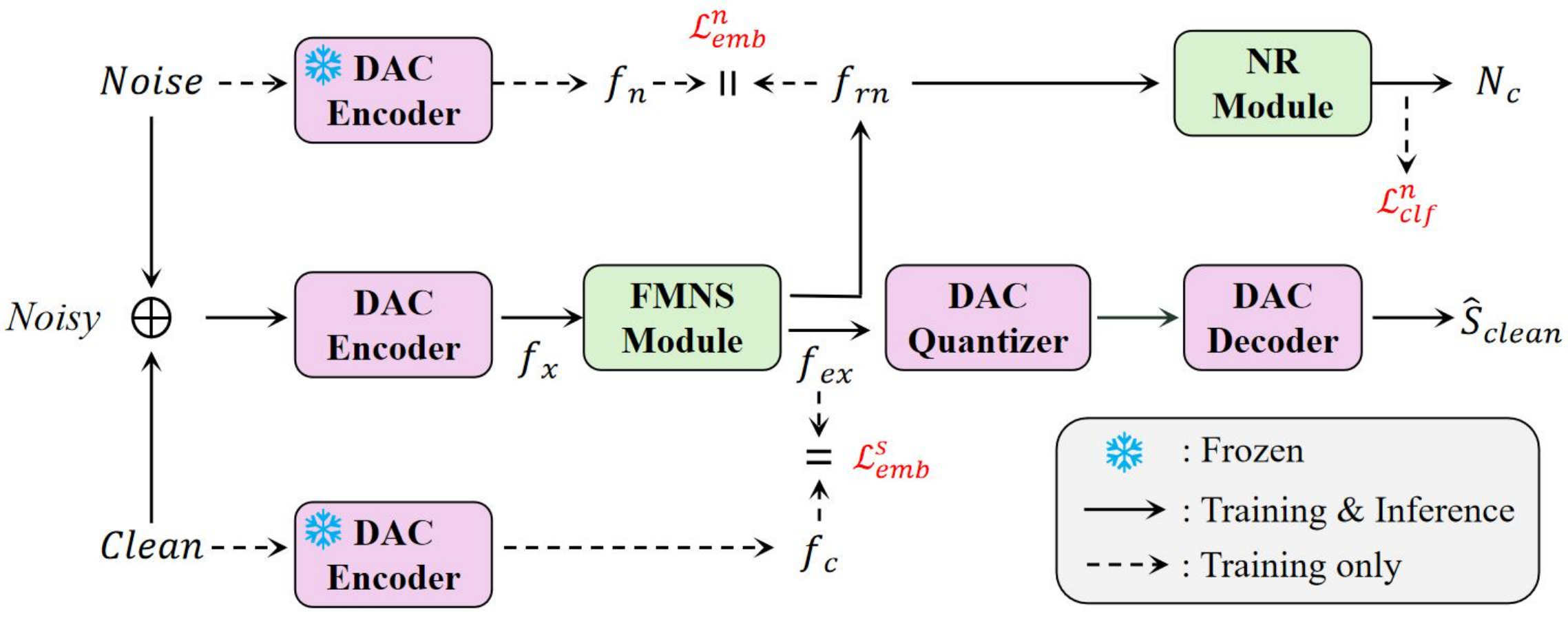}
    }
\end{minipage}%
\hfill
\begin{minipage}{0.36\textwidth}
    \centering
    \subfigure[The focal mask noise separation (FMNS) module.]{
        \includegraphics[width=\linewidth]{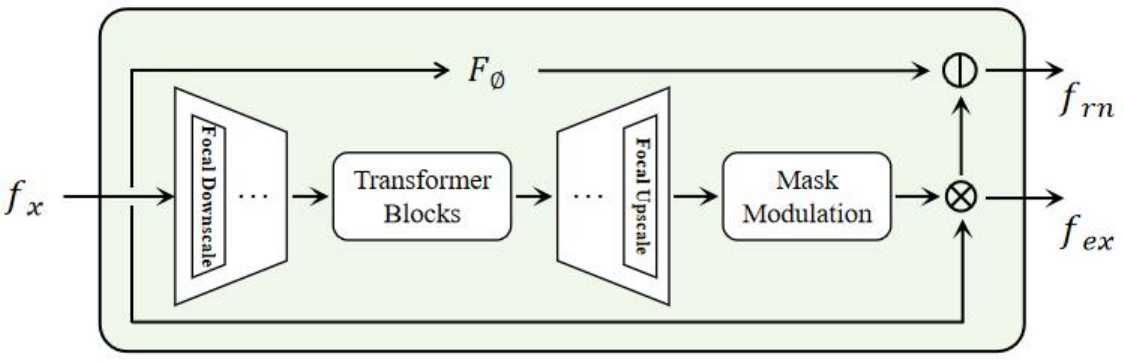}
        \label{subfig:tace}
    }\\
    \subfigure[The noise recognition (NR) module.]{
        \includegraphics[width=\linewidth]{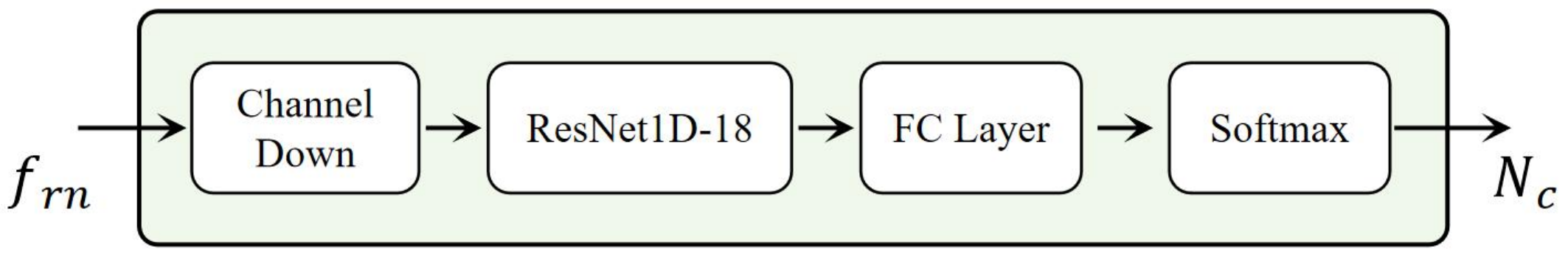}
        \label{subfig:other}
    }
\end{minipage}
\caption{\small The overview of FocalSE. The $f_n$ represents ground-truth noise embedding, $f_{rn}$ represents the reconstructed noise embedding, $f_x$ represents the noisy embedding, $f_c$ represents the clean embedding, $f_{ex}$ represents the enhanced embedding, $N_c$ represents the predicted class probability, and $\hat{S}_{clean}$ represents the reconstructed speech signal.}
\label{fig:flowchart}
\vspace{-0.5cm}
\end{figure*}

\section{Method}
\subsection{Model Architecture}
The proposed FocalSE \footnote{https://github.com/shaokai1209/FocalSE} is a NSC-based SE method that performs feature denoising, noise feature separation and noise recognition in the continuous embedding space of NSC, enabling NSC to adapt to noisy environments. The framework of FocalSE is shown in Figure \ref{fig:flowchart}(a). Key components of FocalSE consist of the focal mask noise separation (FMNS) module and the noise recognition (NR) module.\par
The FMNS module adopts a dual-branch structure, as illustrated in Figure \ref{fig:flowchart}(b). The lower branch explores feature mask based on the focal modulation mechanism and stacks Transformer \cite{vaswani2017attention} blocks in the low-dimensional compressed space, so as to extract enhanced embeddings $f_{ex}$ from noisy embeddings $f_{x}$. The upper branch employs the SEMamba \cite{WOS:001440556800041} $F_{\varnothing}$ to filter the $f_{x}$. The filtered results are then subtracted by $f_{ex}$ to separate the noise embedding features $f_{rn}$. The structure of the NR module is illustrated in Figure \ref{fig:flowchart}(c). It first performs downsampling on the feature channels, then introduces a ResNet1D-18 \cite{cai2023waveform} to extract discriminative features from the $f_{rn}$, and finally obtains noise classification results $N_c$ via a fully connected layer and a softmax layer.\par
\subsection{Training Strategy}
In this paper, we select DAC \cite{kumar2023high} as the basic NSC architecture. The proposed FocalSE performs in the continuous embedding space of DAC, and its training strategy comprises two stages, pre-training and noisy adaptation.\\
\textbf{Pre-training stage.} First, we train the DAC model on clean speech data under given bitrate configurations, enabling it to achieve fundamental signal reconstruction capability.\\
\textbf{Noisy adaptation stage.} This stage integrates the FocalSE into the DAC architecture from the previous stage for feature denoising and reconstructing speech signals $\hat{S}_{clean}$. All model weights except those associated with FocalSE are initialized using pre-trained DAC weights, followed by fine-tuning on the noisy speech dataset. We freeze the weights of the pre-trained DAC encoder, use it as a feature extractor to extract the ground-truth noise embedding $f_n$ and clean embedding $f_c$, and guide FocalSE to perform feature denoising and noise separation.
\subsection{Focal Mask and Noise Separation}
The FMNS module aims to exploit focal mask to extract enhanced embeddings $f_{ex}$ from noisy embeddings $f_{x}$. It further separates noise embeddings $f_{rn}$ to improve the feature denoising capability. The obtained $f_{ex}$ is fed into the quantizer and decoder for speech reconstruction.\par
We introduce the focal downscaling module \cite{della2025focalcodec} to performs feature denoising, which combines the downscaling operation with the focal modulation mechanism. Specifically, it first aggregates global contextual information and then modulates local interactions based on this aggregated representation, enabling it to capture the local variation information of speech signals during the feature compression process. For the decompression process, the downscaling layers are replaced with upscaling layers. On this basis, we stack four Transformer blocks in the low-dimensional compressed space to capture contextual information throughout the compression and decompression processes, thereby facilitating effective feature transition. In this way, FocalSE exploits the global contextual information and local mutual information. We further modulate these information weights with a learnable Sigmoid \cite{luo2019conv} to obtain the focal mask, which is then applied to denoise $f_{x}$ and ultimately generate $f_{ex}$. The calculation process is detailed as follow:
\begin{equation}
\begin{aligned}
f_{ex} = \sigma(M_{focal}(f_{x}))\odot f_{x}
\label{eq:focal_mask}
\end{aligned}
\end{equation}
where $\sigma(\cdot)$ represents the learnable Sigmoid for mask modulation, $M_{focal}(\cdot)$ represents the operation of focal mask. We design and minimize the $\mathrm{L1}$ loss to align the enhanced embeddings with the clean embeddings, as follow:
\begin{equation}
\begin{aligned}
\mathcal{L}^{s}_{emb} = \Vert f_{ex}-f_{c} \Vert_{1}
\label{eq:enc_loss}
\end{aligned}
\end{equation}

We expect the focal mask to simultaneously achieve feature denoising and noise separation with synergistic effects. Therefore, we further separate noise based on the $f_{ex}$. Considering that there always exists a certain deviation between the enhanced result $f_{ex}$ and $f_{c}$, we first introduce SEMamba $F_{\varnothing}(\cdot)$ to filter $f_{x}$, and then subtract $f_{ex}$ to reconstruct the noise embedding $f_{rn}$. The calculation formula is written as follows:
\begin{equation}
\begin{aligned}
f_{rn} = F_{\varnothing}(f_{x}) - f_{ex}
\label{eq:noise_sep}
\end{aligned}
\end{equation}
To align the reconstructed noise embedding $f_{rn}$ with the ground-truth noise embedding $ f_{n}$, we design and minimize the following loss:
\begin{equation}
\begin{aligned}
\mathcal{L}^{n}_{emb} = \Vert f_{rn} - f_{n} \Vert_{1}
\label{eq:noise_loss}
\end{aligned}
\end{equation}
\subsection{Noise Recognition and Training Loss}
To further improve the noise separation performance, we develop a noise classification $C_n(\cdot)$ based on the ResNet1D-18. Specifically, we first scale down the feature channels of the separated noise embedding $f_{rn}$ to 256. Then, we use the 1D-convolution-based ResNet18 to learn its discriminative features. Finally, we obtain the class probability $N_c$ through a fully connected layer and a softmax layer. The calculation of the noise recognition loss is given as follows:
\begin{equation}
\begin{aligned}
\mathcal{L}^{n}_{clf} = CrossEntropy(C_n(f_{rn}),y)
\label{eq:nr_loss}
\end{aligned}
\end{equation}
where $y$ represents the true label vector of noise samples.\par
The final training loss of FocalSE can be written as:
\begin{equation}
\begin{aligned}
\mathcal{L}_{final} =\alpha \mathcal{L}^{s}_{emb} + \beta\mathcal{L}^{n}_{emb} + \gamma\mathcal{L}^{n}_{clf} + \mathcal{L}_{D}
\label{eq:final_loss}
\end{aligned}
\end{equation}
where $\alpha$, $\beta$, and $\gamma$ are trade-off parameters, $\mathcal{L}_{D}$ is the loss of discriminators in DAC.
\begin{table*}[ht]
 \centering
\caption{\small Objective metrics of various models under different low-SNR and low-bitrate conditions, the best performance results are shown in boldface, and the second-best results are marked with underline.}
 \label{tab:accuracy_low_feature}
\renewcommand\arraystretch{1.0}
\setlength{\tabcolsep}{3.8pt}
\scalebox{0.80}{
\begin{tabular}{@{}l c c c c c c c c c c c c c c c c c c @{}}
 \hline
\multirow{2}{*}{Models} &
\multirow{2}{*}{\makecell{Bitrate\\(kbps)$\downarrow$}} &
\multirow{2}{*}{\makecell{Params\\(M)}} &
\multicolumn{3}{c|}{-5 dB} &
\multicolumn{3}{c|}{0 dB}&
\multicolumn{3}{c|}{5 dB}&
\multicolumn{3}{c}{10 dB} \\
\cline{4-6}\cline{7-9}\cline{10-12}\cline{13-15}
 & & &
PESQ $\uparrow$ & STOI $\uparrow$ & \multicolumn{1}{c|}{SI-SDR $\uparrow$} & PESQ $\uparrow$ & STOI $\uparrow$ & \multicolumn{1}{c|}{SI-SDR $\uparrow$} & PESQ $\uparrow$ & STOI $\uparrow$ & \multicolumn{1}{c|}{SI-SDR $\uparrow$}& PESQ $\uparrow$ & STOI $\uparrow$ & SI-SDR $\uparrow$\\
 \hline\hline
DAC \cite{kumar2023high} & 6.00 & 77 & 1.215 & 0.738 & -5.193 & 1.314 & 0.809 & -0.835 & 1.496 & 0.867 & 2.992 & 1.773 & 0.911 & 5.817 \\
SECE \cite{li2025speech} & 6.00 & 196 & 1.955 & 0.810 & 4.294 & 2.281 & 0.908 & 6.010 & 2.573 & 0.919 & 7.001 & 2.806 & 0.937 & 7.191 \\
FD-CBR \cite{chae2025towards} & 6.00 & 83 & 1.975 & 0.871 & 4.516 & 2.305 & 0.911 & 6.178 & 2.589 & 0.930 & 7.193 & 2.839 & 0.942 & 7.646\\
FocalSE$^{-NR/ND}$ & 6.00 & 152 & 1.988 & 0.875 & 4.816 & 2.331 & 0.917 & 6.301 & 2.611 & 0.937 & 7.301 & 2.860 & 0.943 & 7.803 \\
FocalSE$^{-NR}$ & 6.00 & 159 & \underline{2.086} & \underline{0.888} & \underline{5.387} & \underline{2.397} & \underline{0.923} & \underline{6.839} & \underline{2.691} & \underline{0.944} & \underline{7.768} & \underline{2.928} & \underline{0.955} & \underline{8.211} \\
FocalSE & 6.00 & 222 & \textbf{2.116} & \textbf{0.892} & \textbf{5.403} & \textbf{2.432} & \textbf{0.926} & \textbf{6.892} & \textbf{2.728} & \textbf{0.945} & \textbf{7.835} & \textbf{2.970} & \textbf{0.957} & \textbf{8.373} \\
\hline
DAC \cite{kumar2023high} & 2.50 & 77 & 1.181 & 0.708 & -6.676 & 1.261 & 0.779 & -2.279 & 1.422 & 0.841 & 1.237 & 1.627 & 0.879 & 2.867 \\
SECE \cite{li2025speech} & 2.50 & 195 & 1.741 & 0.803 & 2.393 & 1.979 & 0.863 & 3.913 & 2.186 & 0.907 & 4.013 & 2.435 & 0.927 & 5.101 \\
FD-CBR \cite{chae2025towards} & 2.50 & 82 & 1.754 & 0.855 & 2.417 & 2.016 & 0.896 & 3.988 & 2.270 & 0.919 & 4.862 & 2.504 & 0.935 & 5.680 \\
FocalSE$^{-NR/ND}$ & 2.50 & 151 & 1.905 & 0.868 & 3.603 & 2.150 & \underline{0.903} & 4.983 & 2.402 & 0.926 & 5.804 & 2.606 & 0.939 & 6.189 \\
FocalSE$^{-NR}$ & 2.50 & 159 & \underline{1.917} & \underline{0.871} & \underline{3.611} & \underline{2.157} & \textbf{0.911} & \underline{5.009} & \underline{2.437} & \underline{0.929} & \underline{5.826} & \underline{2.615} & \underline{0.940} & \underline{6.274} \\
FocalSE & 2.50 & 221 & \textbf{1.932} & \textbf{0.874} & \textbf{3.635} & \textbf{2.190} & \textbf{0.911} & \textbf{5.016} & \textbf{2.465} & \textbf{0.932} & \textbf{5.879} & \textbf{2.682} & \textbf{0.944} & \textbf{6.316} \\
\hline
\end{tabular}}
\vspace{-0.2cm}
\end{table*}
\section{Experiments}
\subsection{Experimental Setup}
\textbf{Baselines.} We select SECE \cite{li2025speech} and FD \cite{chae2025towards} as the baselines for the proposed FocalSE, as these methods both perform feature denoising in the continuous embedding space of DAC. In particular, we employ the reconstruction results of DAC \cite{kumar2023high} under noisy environments as the reference, to clearly benchmark the denoising gains achieved by these NSC-based SE models.\\
\textbf{Datasets.} Our experiments are conducted on LibriTTS \cite{zen2019libritts} and ESC-50 \cite{piczak2015esc}. LibriTTS is a public English speech corpus that collects 585 hours of speech data from 2,456 speakers, with a sampling rate of 24 kHz. ESC-50 is a public environmental sound dataset containing labeled information for 2,000 short audio clips, covering 50 categories of common sound events and with a sampling rate of 44.1 kHz. In our experiments, all samples are uniformly down-sampled to 16 kHz.\par
We select the train-clean-100 and train-clean-360 subsets of LibriTTS as the clean training dataset with 245.1 hours, and the test-clean-100 subset as the clean testing dataset with 8.6 hours. For noisy data, 8/10 of the noise samples from each category of the ESC-50 are randomly selected and added to the clean training dataset, forming the noisy training dataset. The remaining noise samples are randomly selected and added to the clean testing dataset, forming the noisy testing dataset. The Signal-to-Noise Ratio (SNR) is randomly selected from [-10, -5, 0, 5, 10, 15, 20] dB.\\
\textbf{Training configures.} During the pre-training stage, the DAC model uses a downsampling factor of 512. The 16-layer codebooks use 12-bit vector quantization to achieve 6.0 kbps, while the 8-layer codebooks use 10-bit vector quantization to achieve 2.5 kbps. We train DAC on the clean training dataset and test it on the noisy testing dataset. For noisy adaptation stage, all models are trained based on the architecture of the pre-trained DAC. For SECE, we follow its public parameters for reproduction, freeze the DAC weights during training, and replace its original signal reconstruction loss with the loss of DAC discriminators to ensure fair comparison. We train all models on four RTX 3090 GPUs for 150 epochs. For FD, we adopt its publicly available constant bitrate RVQ (FD-CBR) version.  For FocalSE, the batch size is 32, the learning rate is 0.0001, $\alpha$ = 0.2, $\beta$ = 0.2, and $\gamma$ = 0.5.\\
\textbf{Evaluation metrics.}
We select three objective metrics, PESQ \cite{2001436696923}, STOI \cite{WOS:000293734500024} and SI-SDR \cite{20193007228769}, to evaluate the quality of speech signals reconstructed by the proposed FocalSE and baselines. For the noise recognition of FocalSE, we calculate and report its classification accuracy (ACC).
\subsection{Experimental Results and Analysis}
\subsubsection{Speech Reconstruction Results}
Table \ref{tab:accuracy_low_feature} presents the speech reconstruction results of different models under the conditions of low-SNR and low-bitrate. Notably, FocalSE$^{-NR/ND}$ represents the FocalSE without the noise feature separation and NR module, and FocalSE$^{-NR}$ represents that without the NR module. From these experimental results, we can draw the following important conclusions.\par
First, compared with the reconstruction results of DAC, NSC-based SE methods including SECE, FD-CBR and the proposed FocalSE significantly improve the speech reconstruction quality under low-bitrates and low-SNR conditions. These results fully demonstrate that the clean embedding extraction strategy based on the continuous embedding space enables NSC to adapt well to noisy environments.\par
Second, compared with SECE that freezes the DAC weights during training, both FD-CBR and FocalSE achieve better performance. The possible reason is that the clean embedding extraction strategy adopted by FD-CBR and FocalSE fine-tunes the pre-trained NSC during training. Theoretically, this fine-tuning strategy is superior to the weight freezing strategy, thus enabling superior speech reconstruction results.\par
Third, experimental results show that the proposed FocalSE compared with baselines achieves the best speech reconstruction performance under various conditions of low-bitrate and low-SNR. In contrast to SECE and FD-CBR, the proposed method exploits focal mask that captures local information regarding the short-term variations of speech signals and global contextual information. Furthermore, FocalSE improves the feature denoising capability of the focal mask by separating noise features and noise recognition. In this way, FocalSE obtains higher evaluation results. However, we can observe that FocalSE has a larger number of parameters compared with other models, among which the NR module accounts for approximately 222 - 159 = 63 M, the FMNS module 159 - 77 = 82 M, and DAC 77 M. In other words, if the noise recognition task is not considered, the inference parameters of FocalSE are 159M.\par
Finally, experimental results show that the speech reconstruction performance of FocalSE$^{-NR/ND}$, FocalSE$^{-NR}$ and FocalSE improves steadily. This indicates that removing any component in the ablation study leads to a performance degradation of FocalSE, which fully demonstrates the effectiveness of the FocalSE strategy.
\begin{figure*}[ht]
\vspace{-0.1cm}
\setlength{\abovecaptionskip}{0.cm}
\setlength{\belowcaptionskip}{-0.cm}
  \centering
     \subfigure[ -5 dB (ACC: 72.61$\%$)]{
   \begin{minipage}{3.8cm}
    \includegraphics[width=3.8cm]{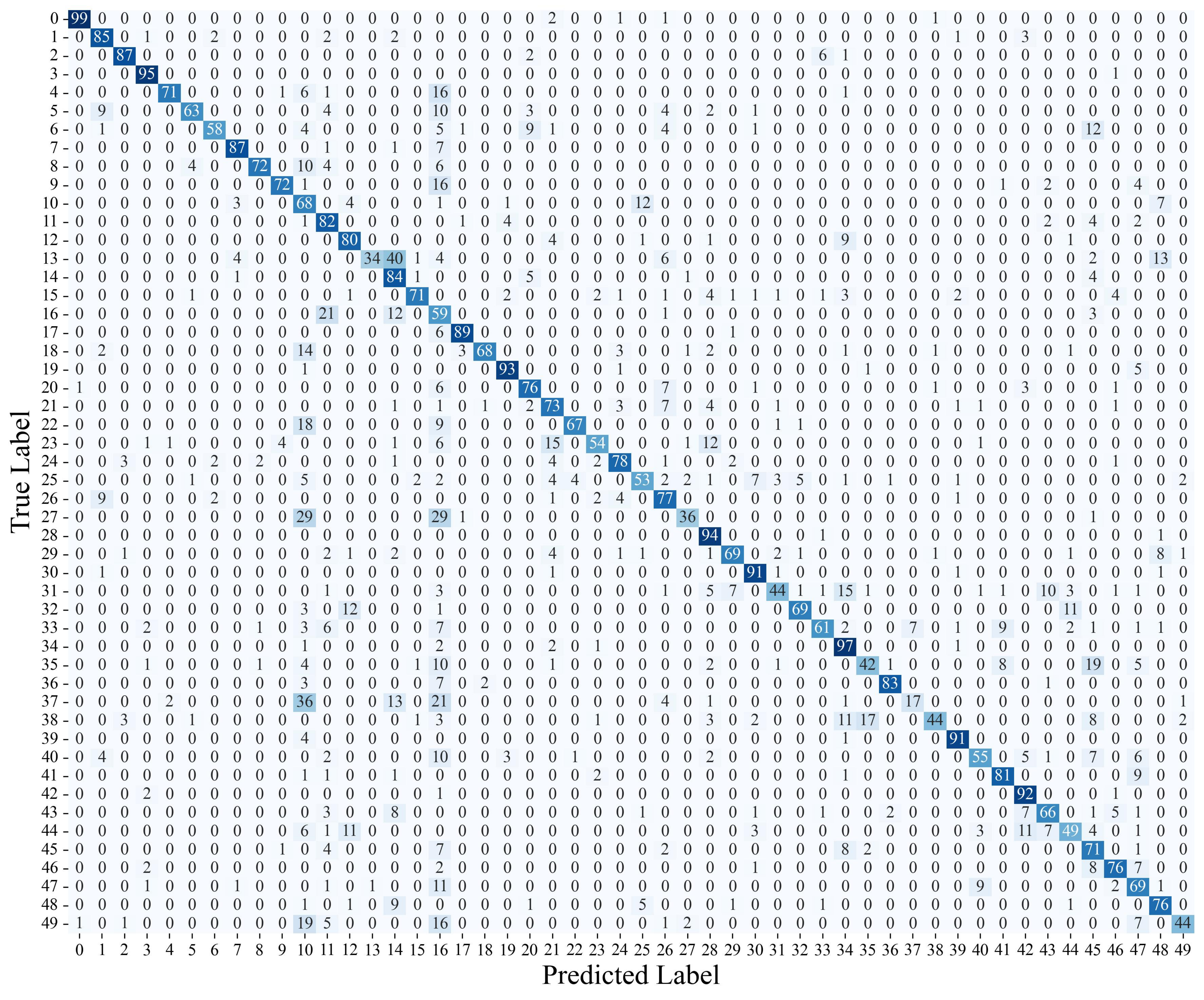}
    \end{minipage}
    }
    \hspace{0.2cm}
    \subfigure[ 0 dB (ACC: 73.52$\%$)]{
   \begin{minipage}{3.8cm}
    \includegraphics[width=3.8cm]{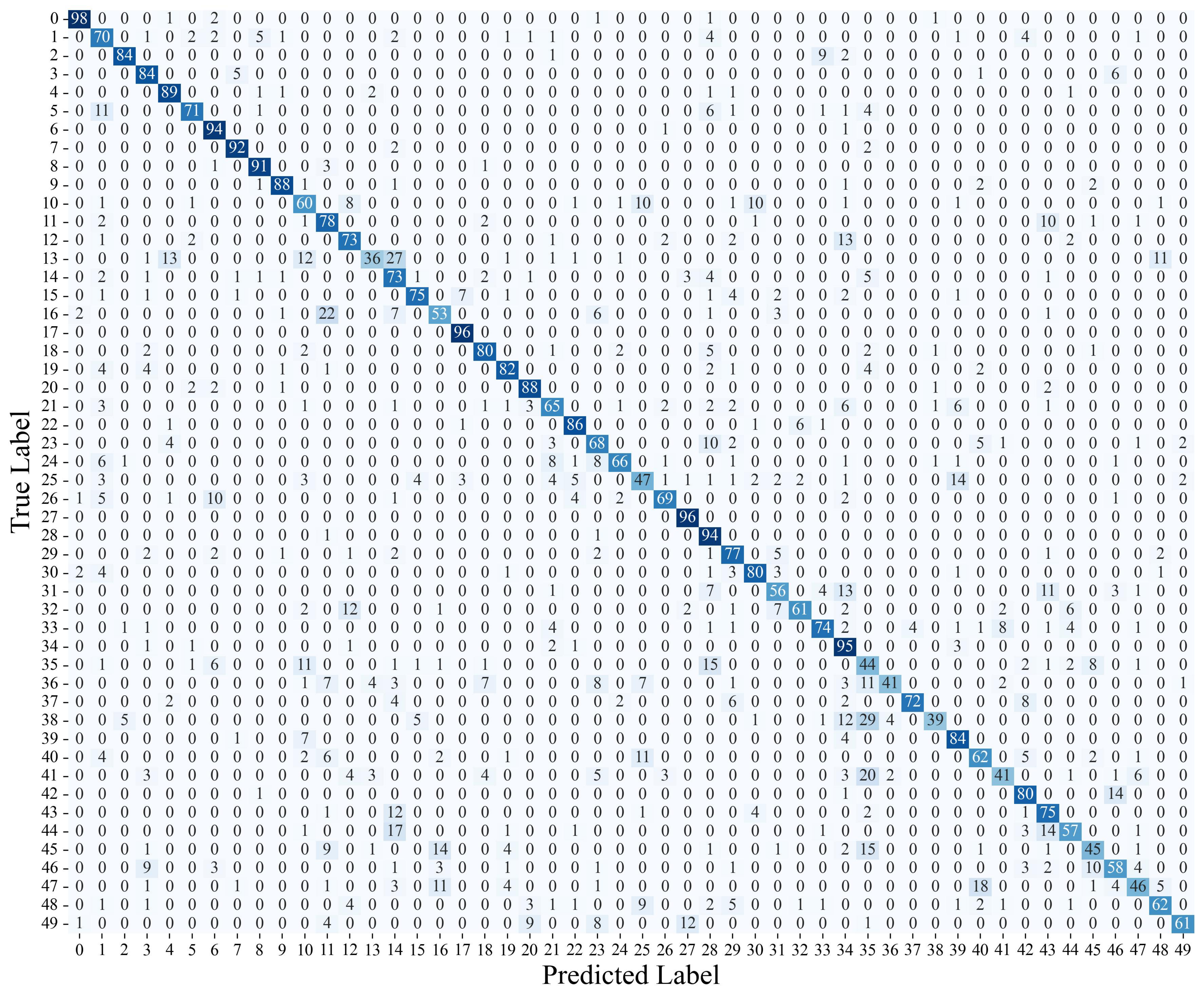}
    \end{minipage}
    }
    \hspace{0.2cm}
    \subfigure[ 5 dB (ACC: 72.67$\%$)]{
   \begin{minipage}{3.8cm}
    \includegraphics[width=3.8cm]{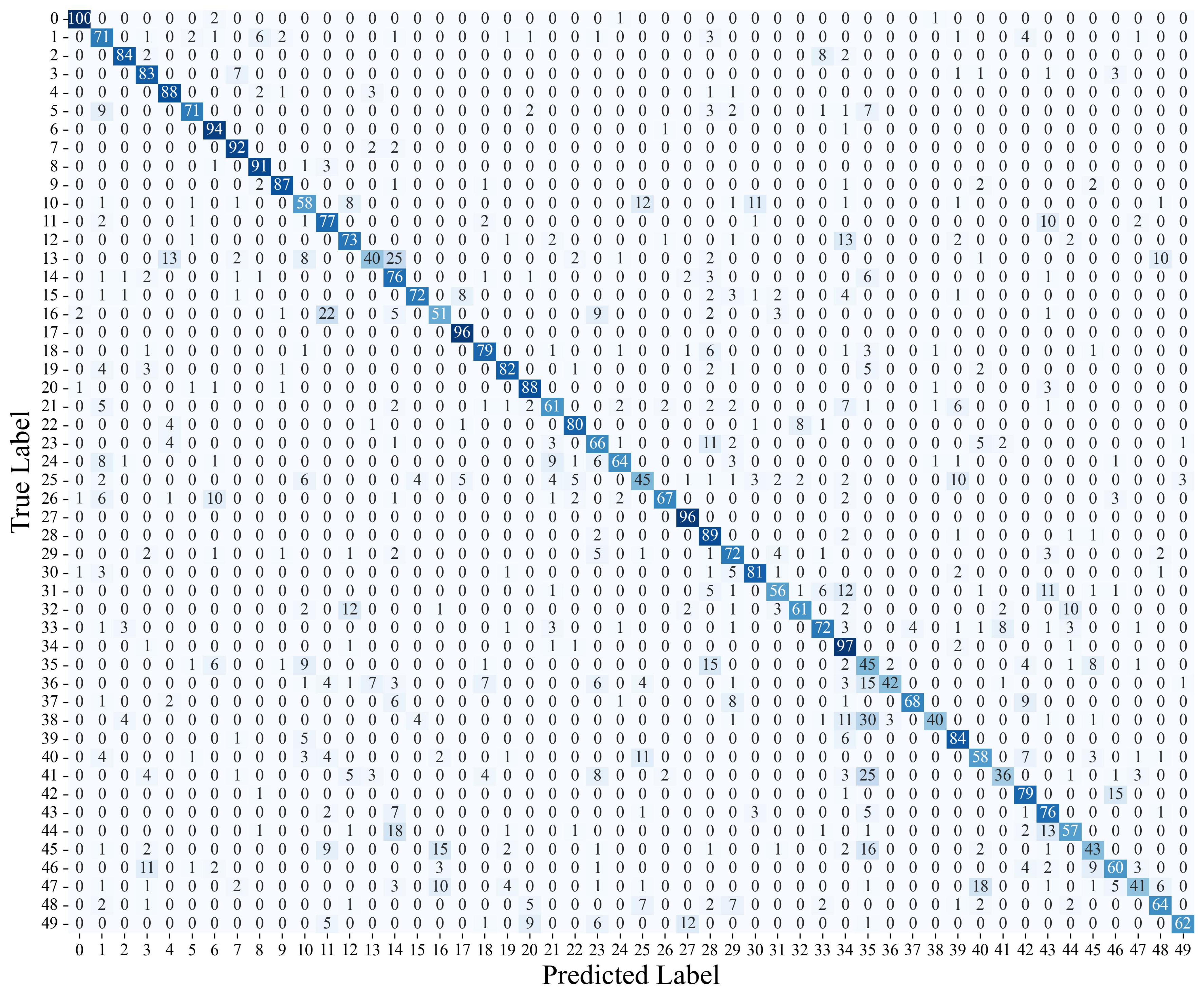}
    \end{minipage}
    }
    \hspace{0.2cm}
    \subfigure[ 10 dB (ACC: 73.02$\%$)]{
   \begin{minipage}{3.8cm}
    \includegraphics[width=3.8cm]{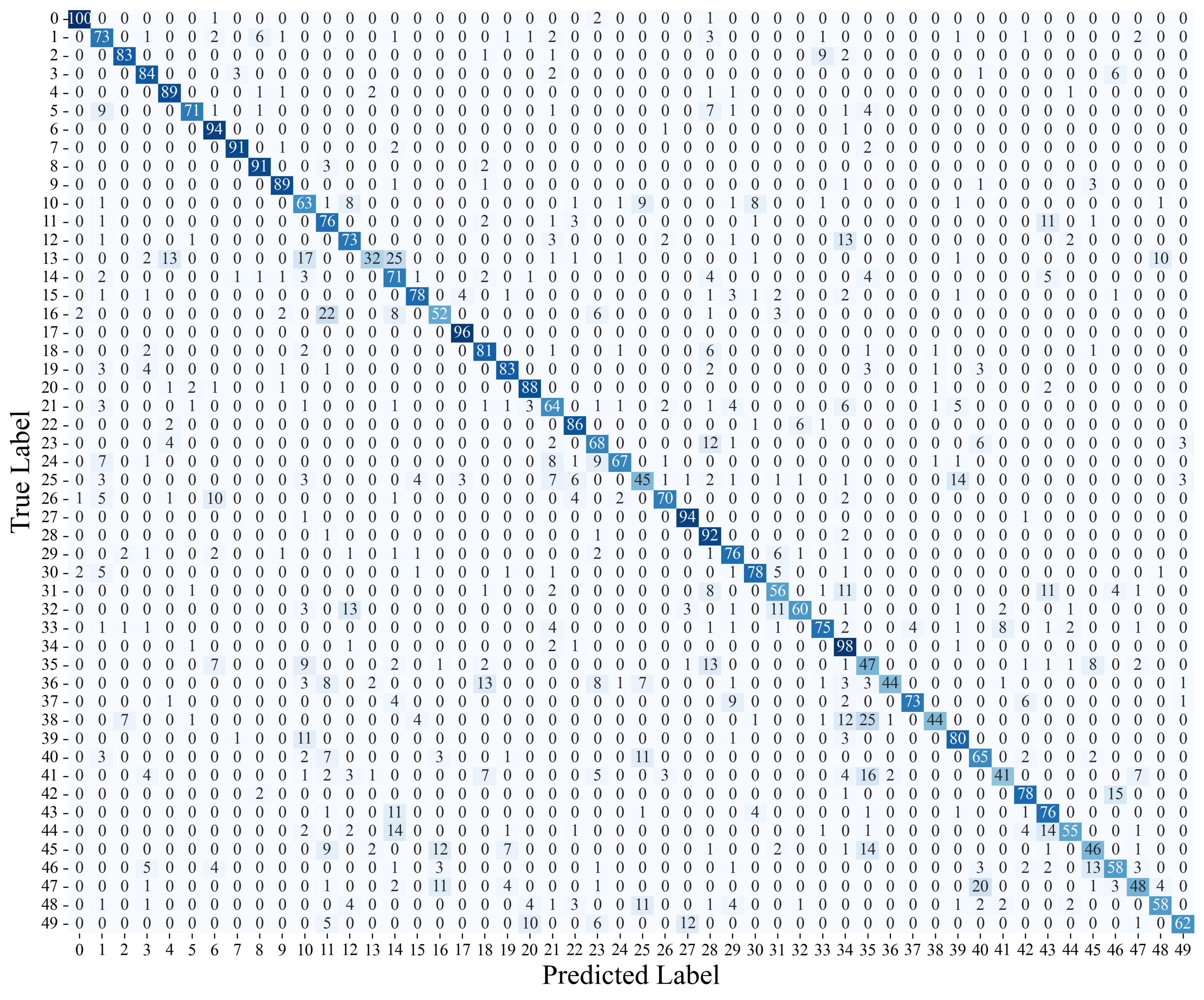}
    \end{minipage}
    }

    \vspace{-0.2cm}

   \subfigure[-5 dB (ACC: 70.89$\%$)]{
   \begin{minipage}{3.8cm}
    \includegraphics[width=3.8cm]{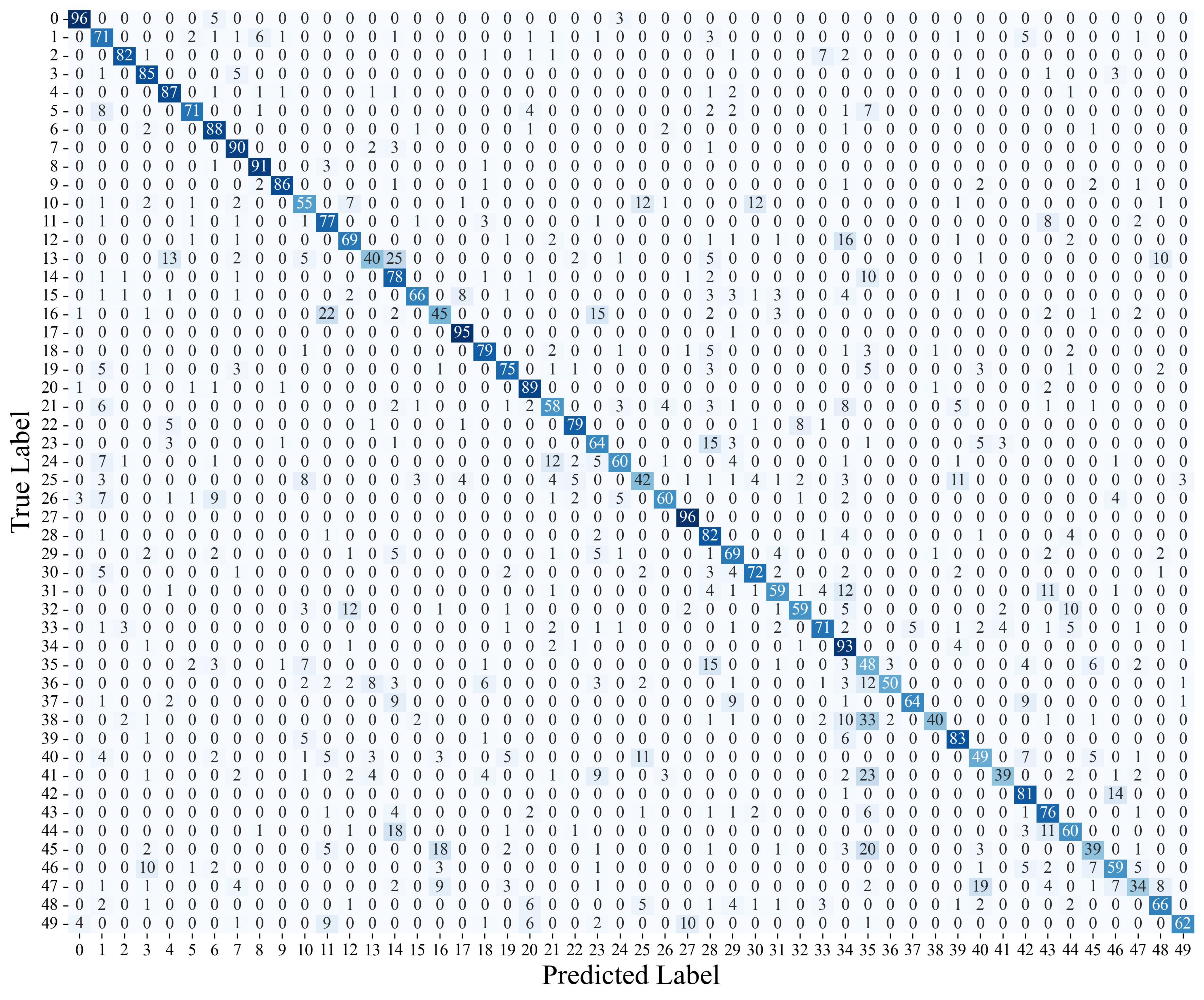}
    \end{minipage}
    }
    \hspace{0.2cm}
    \subfigure[0 dB (ACC: 71.95$\%$)]{
   \begin{minipage}{3.8cm}
    \includegraphics[width=3.8cm]{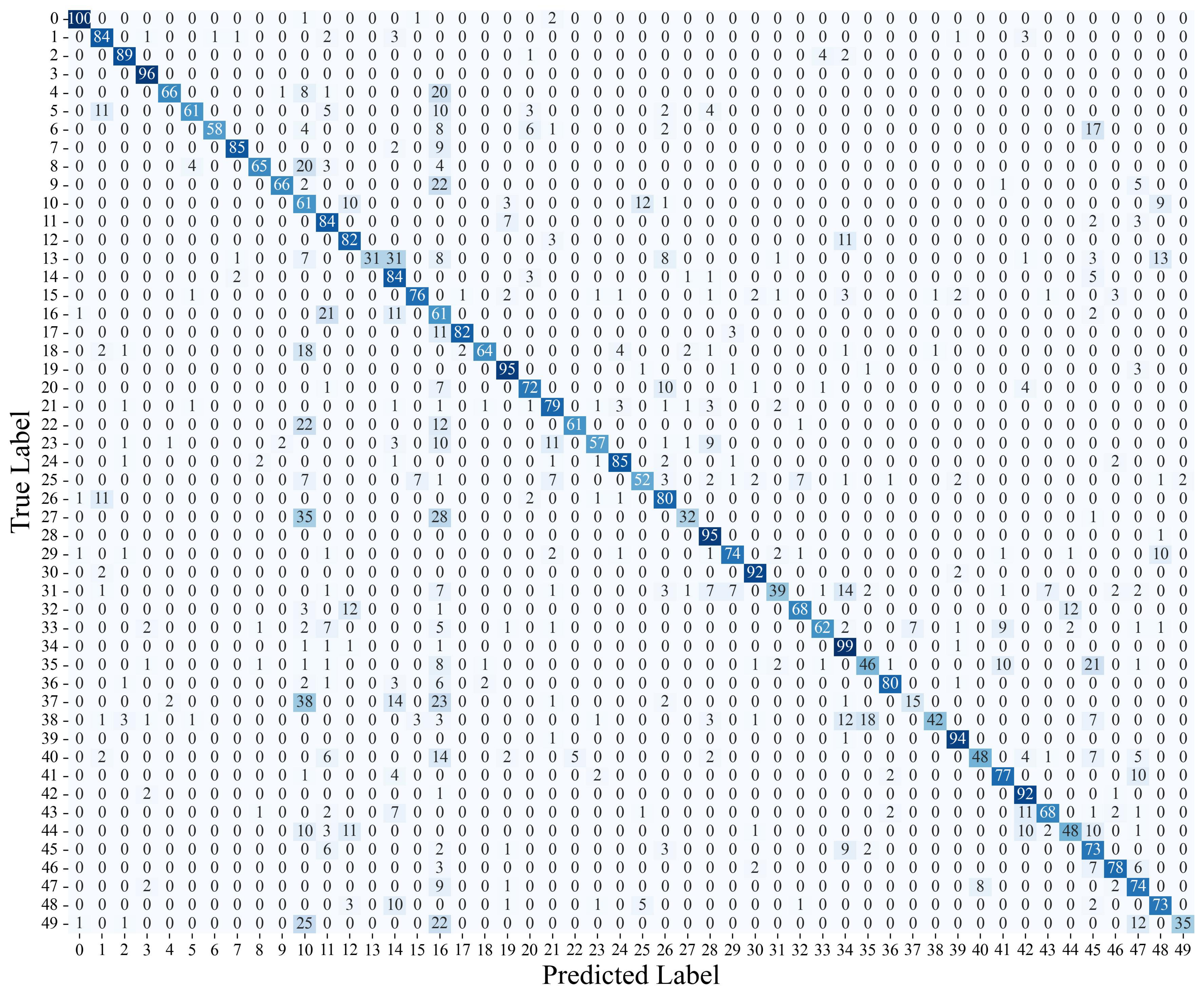}
    \end{minipage}
    }
    \hspace{0.2cm}
    \subfigure[5 dB (ACC: 72.63$\%$)]{
   \begin{minipage}{3.8cm}
    \includegraphics[width=3.8cm]{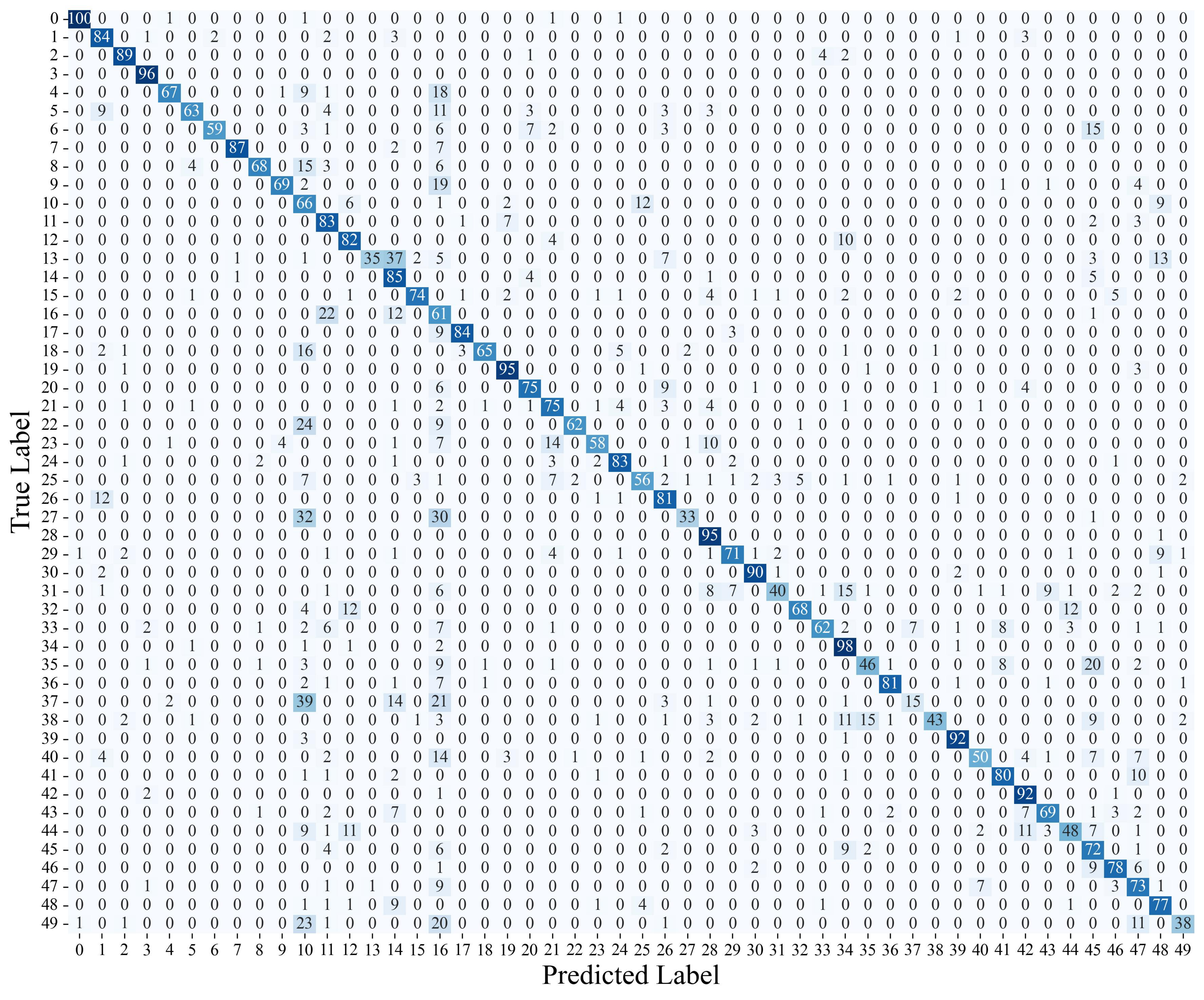}
    \end{minipage}
    }
    \hspace{0.2cm}
    \subfigure[ 10 dB (ACC: 71.10$\%$)]{
   \begin{minipage}{3.8cm}
    \includegraphics[width=3.8cm]{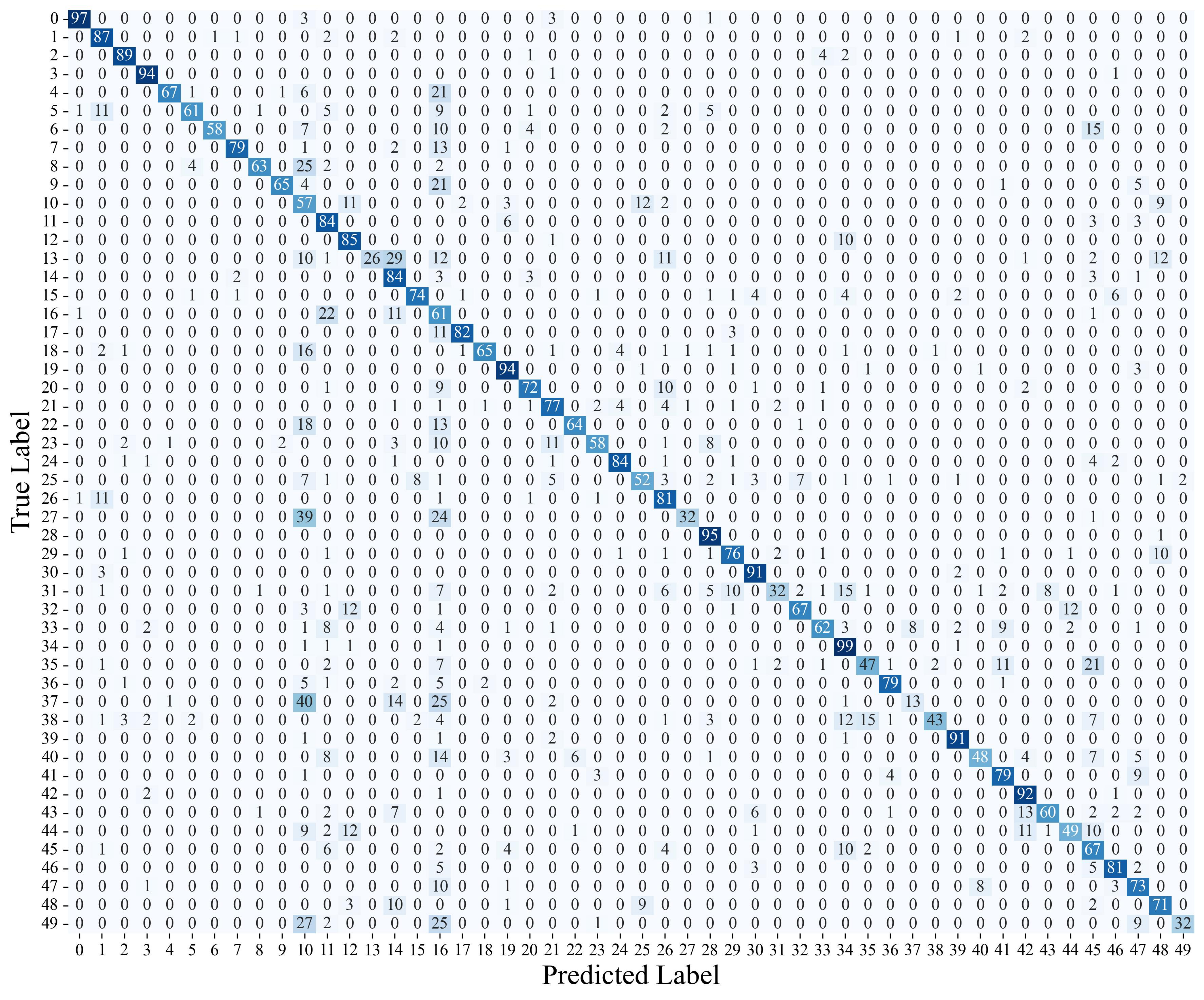}
    \end{minipage}
    }
  \caption{\small Confusion matrices and accuracy (ACC) of FocalSE for noise recognition under different SNR and bitrates, with the upper half for 6.0 kbps and the lower half for 2.5 kbps.}
  \label{fig:confusion}
\vspace{-0.3cm}
\end{figure*}
\subsubsection{Noise Recognition Results}
We present the noise recognition accuracy and confusion matrices of FocalSE in Figure \ref{fig:confusion}. From these results, we can draw the following important conclusions.\par
First, as can be seen, FocalSE achieves accuracy of over 70$\%$ in fine-grained noise classification under different conditions of low-bitrate and low-SNR. It can also be seen from the corresponding confusion matrices that most noise categories exhibit good discriminability, which fully demonstrates the effectiveness of the noise features separated by FocalSE.\par
Secondly, we find that the noise recognition accuracy at 6.0 kbps is mostly higher than that at 2.5 kbps. Besides, the noise classification task under low SNR conditions is more challenging, as decreasing SNR may distort the original energy distribution of the speech signals and thus reduce inter-class differences. More detials are described in \cite{liu2023sound}.
\begin{figure}[t]
\centering
\setlength{\abovedisplayskip}{1pt}
\setlength{\belowdisplayskip}{1pt}
\subfigure[Noisy]{
   \begin{minipage}{3.4cm}
    \includegraphics[width=3.4cm]{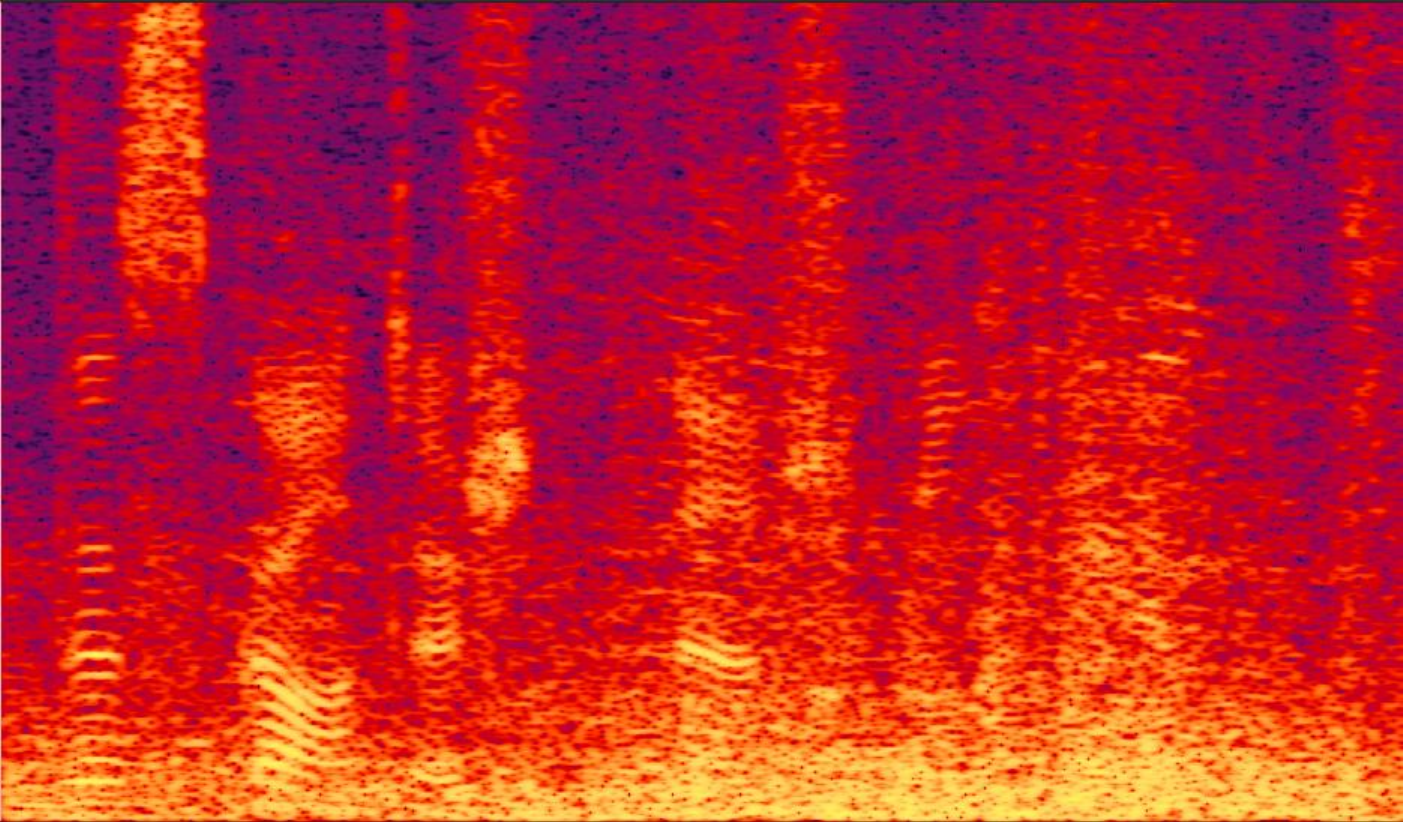}
   \end{minipage}
}
\hspace{0.5cm}
\subfigure[Clean target]{
   \begin{minipage}{3.4cm}
    \includegraphics[width=3.4cm]{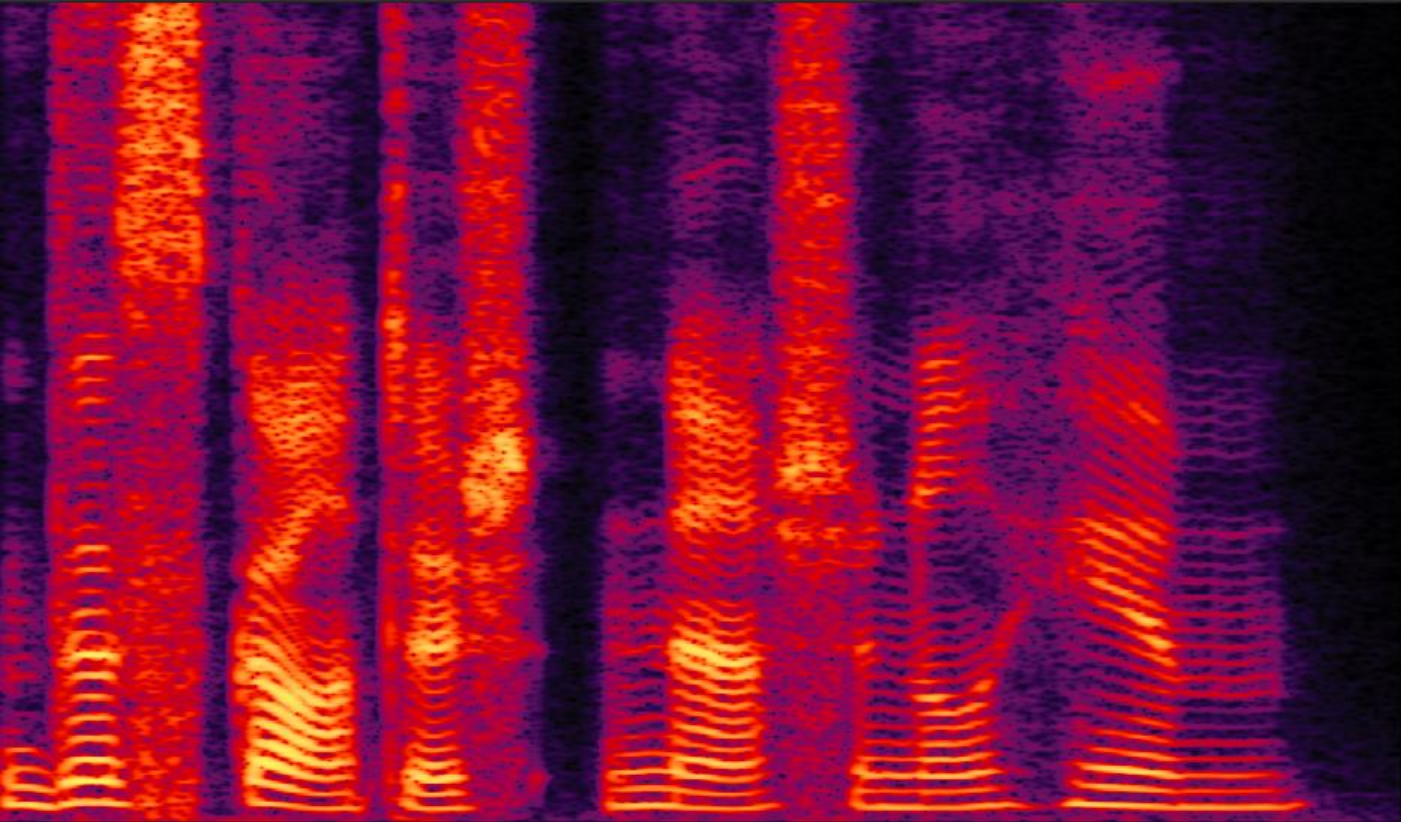}
   \end{minipage}
}

\vspace{-0.25cm}

\subfigure[FD-CBR under 6.0 kbps]{
   \begin{minipage}{3.4cm}
    \includegraphics[width=3.4cm]{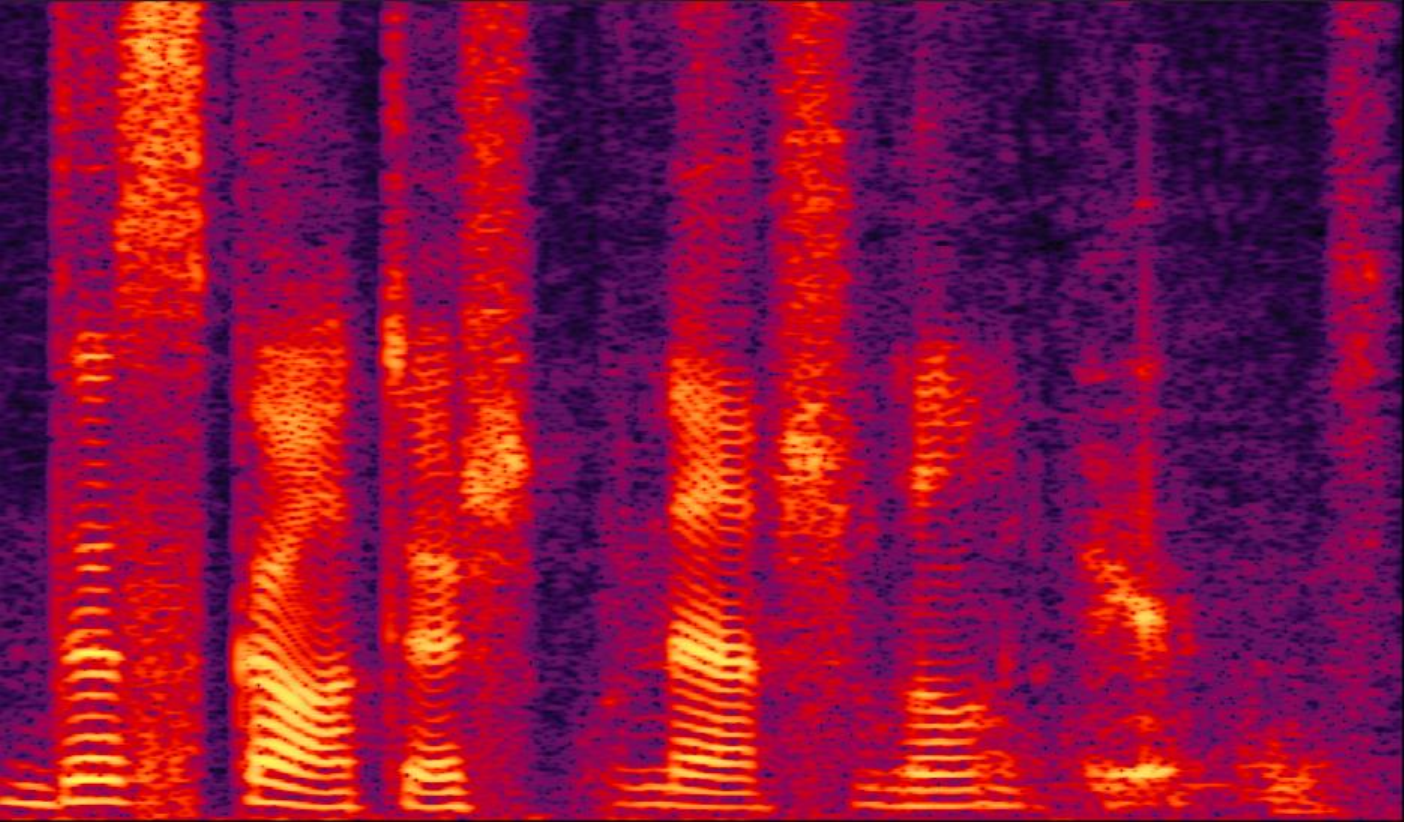}
   \end{minipage}
 }
\hspace{0.5cm}
\subfigure[FocalSE under 6.0 kbps]{
   \begin{minipage}{3.4cm}
    \includegraphics[width=3.4cm]{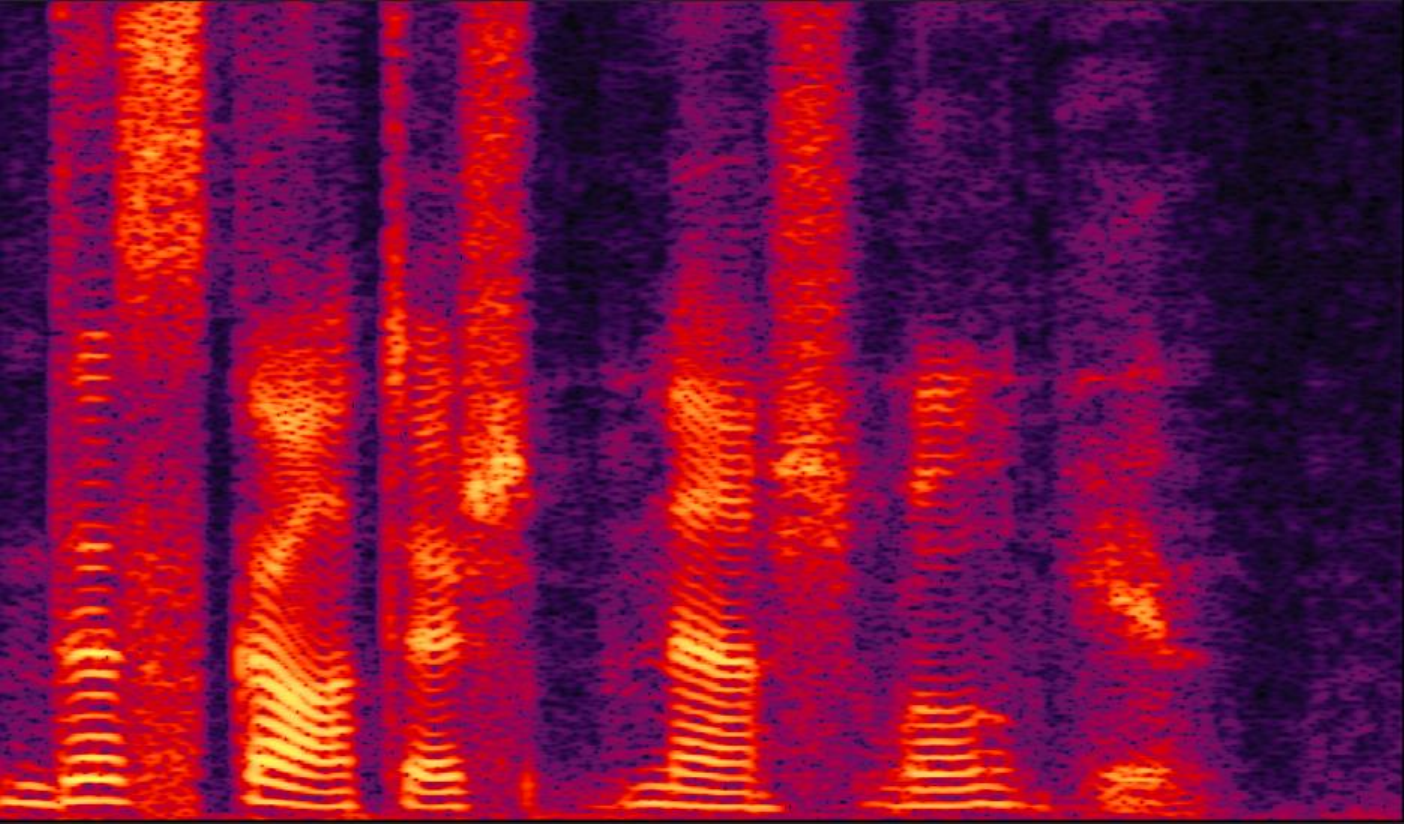}
   \end{minipage}
}

\vspace{-0.25cm}

\subfigure[FD-CBR under 2.5 kbps]{
   \begin{minipage}{3.4cm}
    \includegraphics[width=3.4cm]{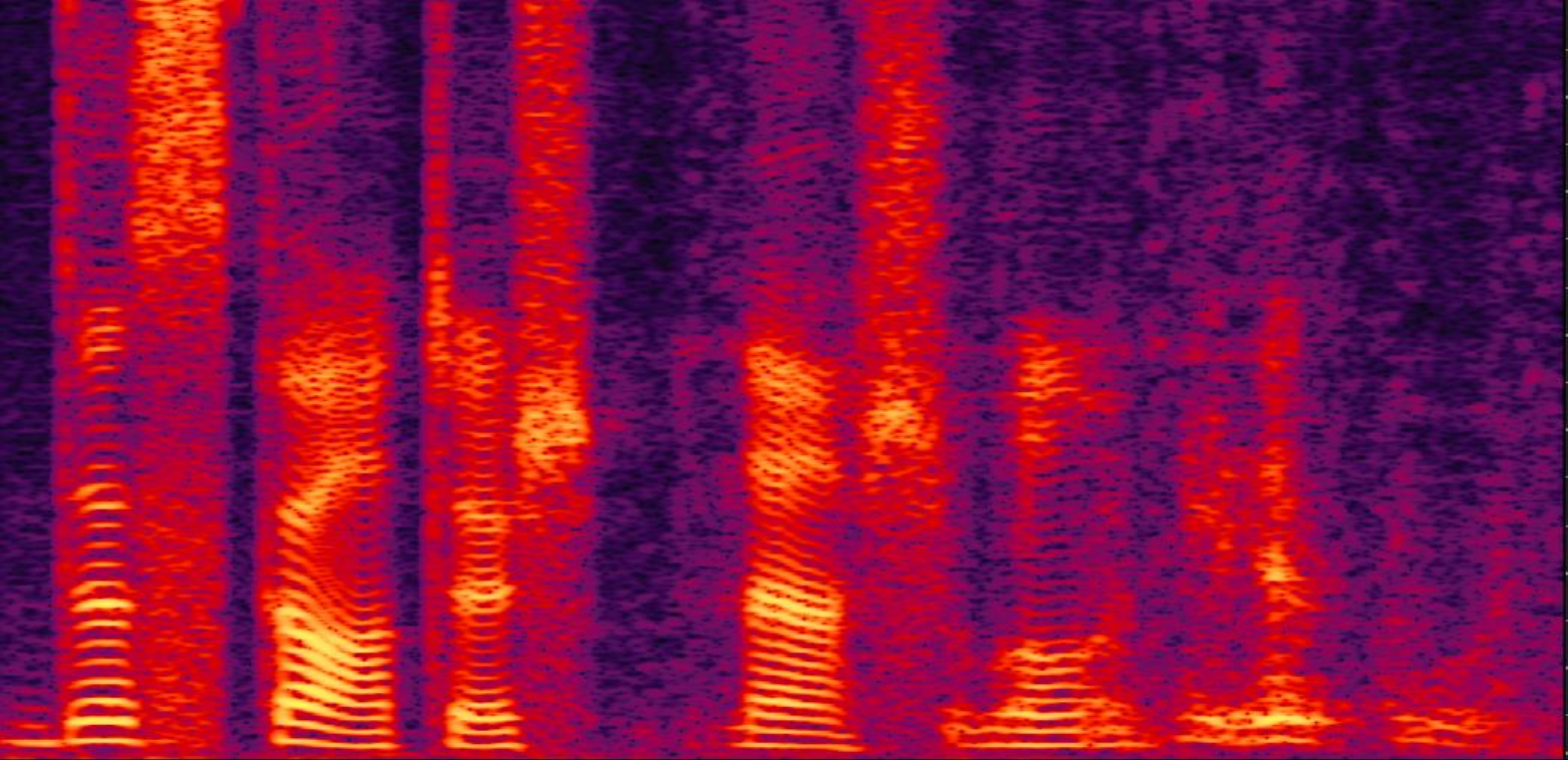}
   \end{minipage}
 }
\hspace{0.5cm}
\subfigure[FocalSE under 2.5 kbps]{
   \begin{minipage}{3.4cm}
    \includegraphics[width=3.4cm]{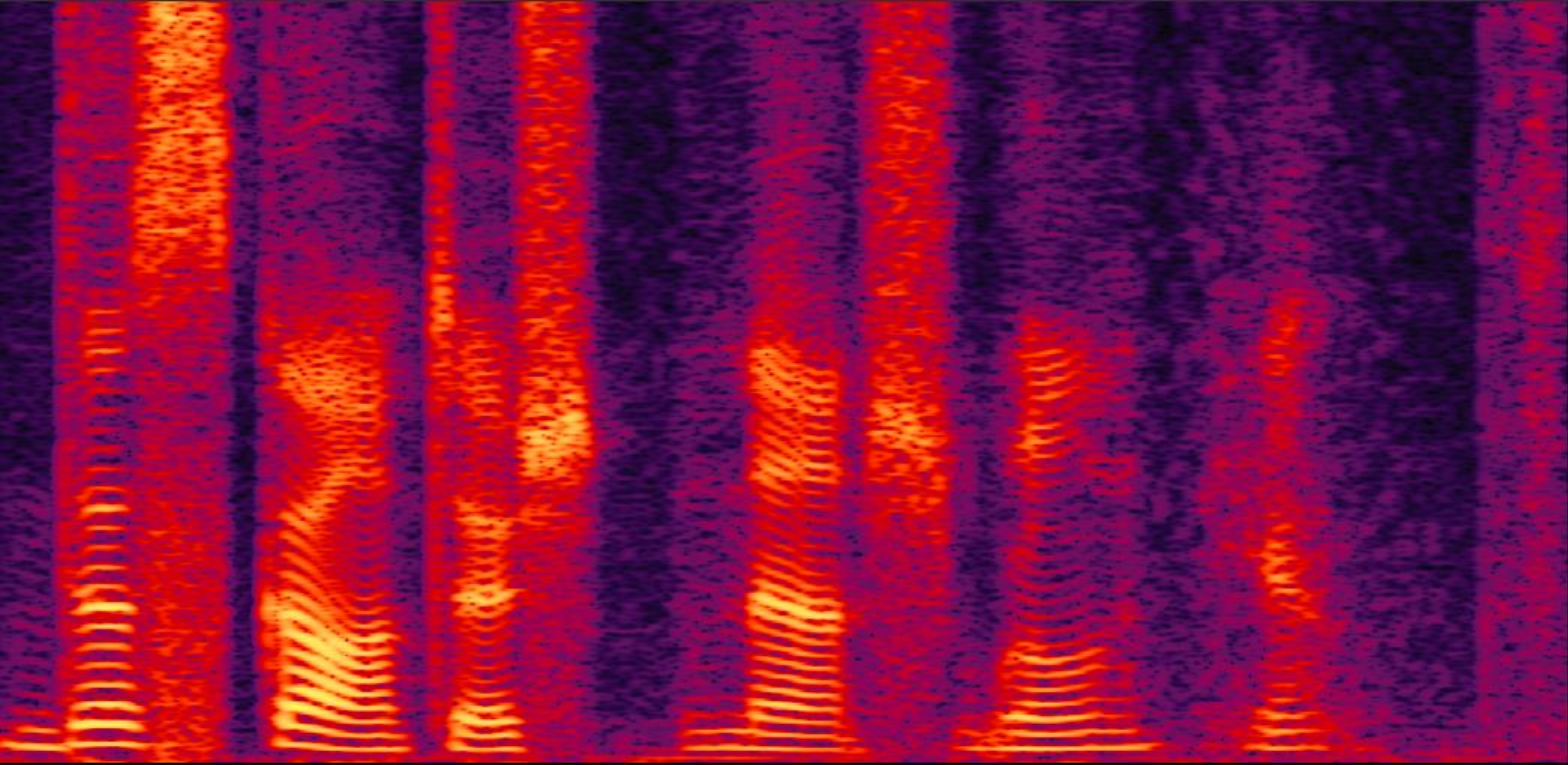}
   \end{minipage}
}

\vspace{-0.25cm}

\caption{\small Spectrograms comparison of different models for restoring speech signal under -5 dB.}
\label{fig:spec}
\vspace{-0.3cm}
\end{figure}
\subsubsection{Spectrograms Comparison for Restoring Speech}
Figure \ref{fig:spec} presents the spectrogram examples of speech reconstructed by different models under the -5 dB condition. Firstly, we can observe that the reconstructed speech by FD-CBR and FocalSE can suppress most of the noise. Secondly, compared with FD-CBR, the energy distribution and harmonic structure of the speech reconstructed by FocalSE are significantly closer to those of the clean target, achieving superior speech reconstruction and denoising results.
\section{Conclusion}
In this paper, we propose a novel NSC-based SE method, named FocalSE, which exploits the focal mask to perform feature denoising, noise feature separation and noise recognition in the continuous embedding space of NSC, enabling NSC to adapt to noisy environments. We conduct extensive experiments and analyses, compared with state-of-the-art methods, the proposed FocalSE achieves superior feature denoising and speech signal reconstruction performance. In the future, we will extend the focal mask to the speech-noise separation under ultra low-bitrate and low-SNR conditions.
\bibliographystyle{IEEEtran}
\newpage
\section{Acknowledgments}
This work is supported by the National Nature Science Foundation of China (No. 62471343, No.62171326).
\section{Generative AI Use Disclosure}
During the preparation of this work, the authors used generative AI tools only for manuscript polishing. After using AI‑assisted tools, the authors reviewed, edited and revised the content as needed, and take full responsibility for the content of this paper.
\bibliography{mybib}

\end{document}